\documentclass{ws06-procs}

\begin{document}

\title{A kinematically decoupled component in NGC4778}

\author{M. Spavone, G. Longo, M. Paolillo}

\address{Universit\'a degli studi di Napoli "Federico II", Dipartimento di Fisica, Via Cinthia 6,
80126,Naples, Italy}

\author{E. Iodice}

\address{INAF, Osservatorio Astronomico di Capodimonte,
Via Moiariello 16, 80131, Naples, Italy}


\maketitle

\abstracts{We present a kinematical and photometrical study of a
member, NGC4778, of the nearest (z=0.0137) compact group: Hickson
62. Our analysis reveals that Hickson 62a, also known as NGC4778,
is an S0 galaxy with kinematical and morphological peculiarities,
both in its central regions (r $< 5''$) and in the outer halo. In
the central regions, the rotation curve shows the existence of a
kinematically decoupled stellar component, offset with respect to
the photometric center. In the outer halo we find an asymmetric
rotation curve and a velocity dispersion profile showing a rise on
the SW side, in direction of the galaxy NGC4776.}

\section{Introduction}
\subsection{Compact Groups}\label{subsec:CG}
Poor groups of galaxies are the most common cosmic structures and
contain a large fraction of the galaxies present in the universe.
At a difference with rich clusters, they span a wide range of
densities, from loose groups, to compact ones.  For this reason,
they are the ideal ground where to test all scenarios for galaxy
formation and evolution and where to pinpoint the details of the
physics controlling galaxy interactions.

Many factors converge in identifying compact groups as good
candidates to be one of the regions where some of this
preprocessing occurs. First, their high spatial density of
luminous matter and small velocity dispersions imply dynamical
lifetimes of the order of a fraction of the Hubble time. This
leading to the possibility that the groups observed in the present
time and in the local universe are second generation objects, just
accreting new members from the loose groups of galaxies in which
almost always they are embedded (\refcite{Vennik93}). Second,
compact groups are numerous and contain a non negligible fraction
of the baryonic matter in the nearby universe
(\refcite{Pildis96}). Therefore, whatever is their ultimate fate,
they are bound to have an impact on the observable properties of
galaxies and cosmic structures.

Our understanding of the dynamical and evolutionary status of
compact groups still presents quite a few gaps. In this respect,
to study the dynamical and evolutive status of a group two
observables are crucial: the detailed kinematics of the individual
galaxies and the structure of the diffuse hot gas halo. In this
work we focus mainly on the peculiar kinematics and on the
photometry of the dominant galaxy NGC4778 (Hickson 62a).

\subsection{HCG62}\label{subsec:HCG62}
Hickson~62 is a quartet of four accordant early type galaxies at a
redshift of 0.0137. The group is dominated by the pair NGC4778
(HCG~62a) and NGC4776 (HCG~62b). NGC4778 formerly classified as an
E3 galaxy, is now classified as an S0 with a bright compact
nucleus, and subsequently discovered to be a low luminosity AGN
(\refcite{Fukazawa01} and \refcite{Coziol98}). Both NGC4776 and
NGC4761 (HCG62c) are classified as peculiar S0's. Finally, NGC4764
(HCG62d) is a faint E2 galaxy.

The compact group is also embedded in a bright X-ray halo which
extends out to 200 kpc, revealing the presence of a deep common
gravitational well and showing the presence of large cavities in
the gaseous halo due to the interaction with the central AGN
(\refcite{Vrtilek02}).

Several studies show that the pair NGC4778/4776 is not interacting
and the kinematical peculiarities observed in NGC4778 are likely
due to an interaction with NGC4761. In fact, the velocity
dispersion and rotation curves of NGC4776 are well behaved and
appear unperturbed, while the velocity dispersion profile of
NGC4778 shows a relatively sharp increase to the SE, suggestive of
the presence of a perturber.

\section{Results}\label{sec:results}
\subsection{Photometry}\label{subsec:phot}
We have used ESO archive CCD images of HCG62 obtained with FORS1
at ESO VLT-UT1 in the Jonson B and R bands. Here we briefly
describe and discuss the main features in the light distribution
of NGC4778. The photometrical analysis shows the presence of a
component with round isophotes for $2''<r<12''$. For $r>R_{e}$,
$\epsilon$ increase and the P.A. changes of about $30^{\circ}$,
suggesting that the isophotes are more crushed and non-coaxial
with the outer one. Moreover, the diskiness has decreasing value
with a minimum at $24''$, corresponding to the maximum values of
$\epsilon$ and P.A. In the nuclear regions the P.A. changes of
about $15-20^{\circ}$, and $a_{4}/a>0$, suggesting the presence of
disky isophotes. The mean B-R color profile of the galaxy shows
the presence of bluer region in the center of NGC4778. The mean
color profile of NGC4778 is consistent with the range of values
typical for early-type galaxies (\refcite{Fukugita95}) and for
spheroidal galaxies in compact groups (\refcite{Zepf91}).

\subsection{Kinematics}\label{subsec:kin}

\begin{figure}[ht]
\begin{center}
\epsfxsize=8cm   
\epsfbox{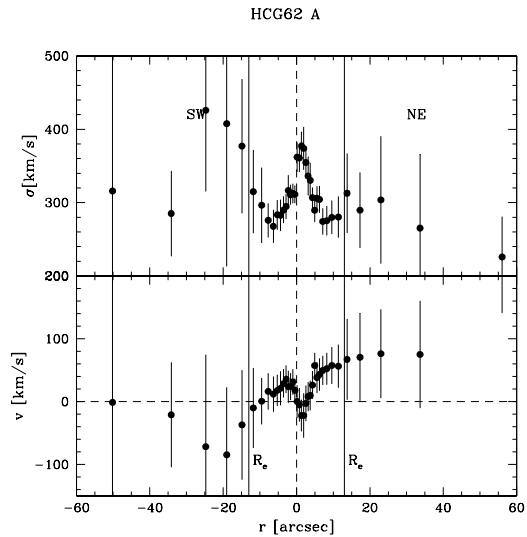}
\centerline{\epsfxsize=3.9in} \caption{Rotation curve and velocity
dispersion profile derived for NGC4778 for the whole galaxy
extension. \label{kin}}
\end{center}
\end{figure}

The \emph{Line-Of-Sight Velocity Distribution} (LOSVD) was then
derived from the continuum-removed spectra using the Fourier
Correlation Quotient (FCQ) method (\refcite{Bender90};
\refcite{Bender94}). The rotation curve and the radial velocity
dispersion profile along the major axis of NGC4778 are shown in
Fig.\ref{kin}. The kinematic profiles reveal two important
features: {\it i)} a counter-rotation in the nuclear region, {\it
ii)} at large galactocentric distances (for $r >2.5''$) the
rotation curve and velocity dispersion are asymmetric with respect
to the galaxy center\footnote{Note that the central value is
chosen in order to obtain the rotation curve symmetric within
$3''$.}.

On the whole, the galaxy presents significant rotation. At large
radii the spectrum is contaminated by the light coming from the
galaxy NGC4761, therefore the value of $200\ km/s$ detected for $r
= 55''$ cannot be due to the typical motion of the stars in
NGC4778. Looking in more detail the nuclear region, for $r\leq
2.5''$ we detect an inversion of the velocity gradient towards the
center, with a maximum value of the velocity of $20-30\ km/s$,
with respect to the outer radii. This inversion within the central
regions with respect to the overall trend, reveals the presence of
a counter-rotating decoupled core. In correspondence with the
inversion of the velocity gradient, the velocity dispersion
profile shows an hint for a local minimum.

\section{Discussion}\label{sec:disc}

We observe an inversion in the velocity profile gradient in the
center, which also correspond to some anomalous photometric
features, such as bluer colors, and twisting in the position angle
of the isophotes. These features strongly suggest the existence of
a small core ($\sim 600\ pc$) kinematically decoupled from the
whole galaxy.

The rotation curve of NGC4778 is not symmetric with respect to the
center, and this is a feature observed in many other compact
groups (\refcite{Bonfanti99}). The simulations performed by Combes
et al.(\refcite{Combes95}) show that the peculiarities observed in
many rotation curves of galaxies belonging to compact groups are
due to intrinsic effects and not to contamination along the line
of sight.

The asymmetry and the shape of the rotation curve and velocity
dispersion profile of NGC4778 do not find correlation with the
photometric features of the galaxy, except for the bluer colors in
the central region. The absence of correlation between the
dynamical and the morphological peculiarities suggests that the
dynamical properties of the HCG galaxies may be due to a minor
merger event. In fact, as showed by Nishiura et
al.(\refcite{Nishiura00}), weak galaxy collisions could not
perturb the galaxy rotation curves, but morphological deformations
could be induced in the outer parts of the galaxy (tidal tails,
bridges etc), while minor mergers could perturb the rotation
curves in the inner regions, especially for gas-poor early-type
galaxies, without causing morphological peculiarities.

\section*{Acknowledgments}
The authors very grateful to R. Saglia. M.S. also wishes to thank
the INAF-Observatory of Capodimonte and N. Napolitano. This work
was funded through a grant from Regione Campania (ex legge 5) and
a MIUR grant.  This work is based on observations made with ESO
Telescopes at the Paranal Observatories under programme ID
$<169.A-0595(C)>$ and $<169.A-0595(D)>$.


\begin{thebibliography}{0}
\bibitem{Bender90} Bender, R., A\&A, {\bf 229}, 441
(1990).
\bibitem{Bender94} Bender, R., Saglia, R.P., and Gerhard, O.E., MNRAS, {\bf 269},
785 (1994).
\bibitem{Bonfanti99} Bonfanti, P., Simien, F., Rampazzo, R. and Prugniel, Ph., 1999, A\&AS, {\bf 139},
483 (1999).
\bibitem{Combes95} Combes, F., Rampazzo, R., Bonfanti, P.P., Prugniel, P. and Sulentic, J.W., A\&A, {\bf 297},
37 (1995).
\bibitem{Coziol98} Coziol, R., Ribeiro, A. L. B., De Carvalho, R. and Capelato, H. V., ApJ, {\bf 493},
563 (1998).
\bibitem{Fukazawa01} Fukazawa, Y., Nakazawa, K., Isobe, N., Ohashi, T. and Kamae, T., ApJ, {\bf 546},
87 (2001).
\bibitem{Fukugita95} Fukugita, M., Shimasaku, K., Ichikawa T., Publications of the astronomical society of the Pacific, {\bf 107},
945 (1995).
\bibitem{Mendes03} Mendes de Oliveira C., Aram, P., Plana, H. and Balkowski, C., AJ, {\bf 126},
2635 (2003).
\bibitem{Nishiura00} Nishiura, S., Shimada, M., Ohyama, Y., Murayama, T. and Taniguchi, Y., AJ, {\bf 120},
1691 (2000).
\bibitem{Pildis96} Pildis, R. A., Evrard, A. E., and Bergman, J. N., AJ, {\bf 112},
378 (1996).
\bibitem{Rubin91} Rubin V.C., Hunter D.A., Ford W.K.Jr., ApJS, {\bf 76}, 153 (1991).
\bibitem{Vennik93} Vennik, J., Richter, G. M. and Longo, G., AN, {\bf 314},
393 (1993).
\bibitem{Vrtilek02} Vrtilek, J., M., Grego, L., David, L. P. et al., APS meeting,
B17.107 (2002).
\bibitem{Zepf91} Zepf, S. E., Whitmore, B. C., Levison, H. F., ApJ, {\bf 383},
524 (1991).
\end{thebibliography}
\end{document}